\documentclass[amsmath,amssymb,
showpacs,twocolumn,
superscriptaddress,
prl,longbibliography]{revtex4-2}
\usepackage{graphicx,bm,color,subfigure}
\usepackage[T1]{fontenc}
\usepackage{mathrsfs}
\usepackage{bm}
\usepackage{amsmath}
\usepackage{amssymb}
\usepackage{graphicx}
\usepackage{esint}
\usepackage{multirow}
\usepackage{float}
\usepackage{array}
\usepackage{makecell}
\usepackage{harpoon}
\usepackage{booktabs}
\usepackage{gensymb}
\usepackage{simplewick}
\usepackage{subfigure}
\usepackage{soul}

\definecolor{ZXBlue}{RGB}{0,125,255}

\newcommand{\be}{\begin{eqnarray}}
\newcommand{\ee}{\end{eqnarray}}

\graphicspath{{pic/}}


\usepackage[colorlinks]{hyperref}

\begin{document}
\title{Classification of Magnetism and Altermagnetism in Quasicrystals}
\author{Zhi-Yan Shao}
\thanks{These authors contributed equally to this work.}
\affiliation{School of Physics, Beijing Institute of Technology, Beijing 100081, China}

\author{Chen Lu}
\thanks{These authors contributed equally to this work.}
\affiliation{School of Physics, Hangzhou Normal University, Hangzhou 311121, China}

\author{Zhiming Pan}
\thanks{These authors contributed equally to this work.}
\affiliation{Department of Physics, Xiamen University, Xiamen 361005, Fujian, China}

\author{Yu-Bo Liu}
\email{yuboliu@itp.ac.cn}
\affiliation{Institute of Theoretical Physics, Chinese Academic of Science, Beijing 100080, China}

\author{Fan Yang}
\email{yangfan\_blg@bit.edu.cn}
\affiliation{School of Physics, Beijing Institute of Technology, Beijing 100081, China}
\date{August 2025}

\begin{abstract}
Altermagnetism (AM), an unconventional magnetic phase characterized by zero net magnetism protected by symmetry(s) other than parity-time ($\mathcal{P}\mathcal{T}$) and a resulting spin-split band, has been studied exclusively in crystalline materials. Here, we extend the framework of AM to quasicrystals (QCs). We start from a comparison between the N\'{e}el state on the square lattice and that on a $D_4$-symmetric Thue-Morse QC, with both belonging to the same $d$-wave irreducible representation (IRRP) of the $D_4$ point group. Consequently, while the former is antiferromagnetism (AFM) protected by the combined $\mathcal{P}\mathcal{T}$ and translational symmetry, the lack of translational symmetry in the latter breaks the $\mathcal{P}\mathcal{T}$ symmetry, and the additional mirror or rotation symmetry protects AM. This example suggests that AM is more common in QCs than in crystals and can be easily explored through a point-group symmetry-based classification. Therefore, we classify magnetic phases in 2D $D_n$-symmetric QCs without spin-orbit coupling, by using IRRPs of $D_n$. Consequently, the identity IRRP represents ferromagnetism, the inversion-odd 1D IRRPs for twice-of-odd $n$ represent AFM, and all the remaining 1D IRRPs represent AM, protected by either mirror or rotation symmetry. We further take the Hubbard model to verify this result in various QCs with different symmetries. Our work highlights the QC as a natural platform where AM is common among magnetic phases. 
\end{abstract}

\maketitle

\paragraph{\textcolor{ZXBlue}{Introduction. ---}} 
Altermagnetism (AM) \cite{vsmejkal2022emerging,vsmejkal2022beyond,mazin2022altermagnetism} has recently emerged as a distinct class of magnetic order ~\cite{naka2019spin,ahn2019antiferromagnetism,hayami2019momentum,vsmejkal2020crystal,yuan2020giant,shao2021spin,mazin2021prediction,ma2021multifunctional,yuan2021prm,PhysRevLett.132.056701,PhysRevLett.133.056401,bai2024altermagnetism,han2024electrical,zhou2025manipulation}, characterized by two key properties: zero net magnetization similar to antiferromagnetism (AFM), and a spin-split electronic band structure characteristic of ferromagnetism (FM).
This duality makes AM a promising candidate for next-generation spintronic applications ~\cite{bai2024altermagnetism,che2024realizing,wu2024valley,ezawa2025third,bai2023efficient,song2025altermagnets,sourounis2025efficient},
enabling both efficient spin-current manipulation and robust magnetic control without the undesirable stray fields associated with FM.


Symmetry-based classification of magnetic states \cite{brinkman1966theory,liu2022spin,xiao2024spin,chen2024enumeration,jiang2024enumeration,yang2020unlocking} is crucial for the exploration of AM. 
From the aspect of symmetry, AM is a magnetic phase without combined parity ($\mathcal{P}$) and time-reversal ($\mathcal{T}$) symmetry, 
while exhibiting zero net magnetism protected by other symmetries. 
AM and other unconventional magnetic orders ~\cite{wu2007fermi,jungwirth2025altermagnetism} are systematically classified using the spin-space group, which incorporates both spatial and spin symmetries in the absence of spin-orbital coupling (SOC) ~\cite{brinkman1966theory,liu2022spin,xiao2024spin,chen2024enumeration,jiang2024enumeration}. 
From a representation theory perspective, AM corresponds to a 1D, non-trivial, inversion-even irreducible representation (IRRP) of the magnetic order parameter \cite{vsmejkal2022emerging,vsmejkal2022beyond,mcclarty2024landua}, which breaks the combined $\mathcal{P}\mathcal{T}$ symmetry.

The material realization of AM is the recent research focus. AM can arise from various mechanisms, including the presence of nonmagnetic atoms ~\cite{fender2025altermagnetism,vsmejkal2022beyond}, spontaneously broken orbital order ~\cite{leeb2024spontaneous,vila2025orbital,meier2025anti}, 
or Pomeranchuk instabilities ~\cite{wu2004dynamic,wu2007fermi,jungwirth2025altermagnetism,qian2025fragile}.
Engineered routes to AM have been proposed in several platforms. For example, twisted van der Waals ferromagnets can host moir\'e superlattices with spin-split bands ~\cite{He2023prl,liu2024twisted,sheng2025ubiquitous}, and inversion-symmetry-broken systems such as ferroelectrics can also exhibit AM~\cite{vsmejkal2024altermagnetic,duan2025antiferroelectric,gu2025ferroelectric,zhu2025two,sun2025proposing,sun2024stacking}. 
Furthermore, external tuning through strain, lattice vacancies, or the formation of spin clusters has been shown to stabilize extrinsic AM~\cite{chakraborty2024strain,belashchenko2025giant,jiang2025strain,zhu2025design,li2025pressure}. In the search for intrinsic AM materials, although many materials have been theoretically predicted to be AM, only a few of them are supported by experimental evidences~\cite{lin2024observation,jiang2025metallic,osumi2024observation,graham2025local,reichlova2024observation,jiang2024discovery,reimers2024direct}, with the confirmed ones even rarer. While previous studies of AM are mostly focused on conventional crystalline materials, here we demonstrate that the quasicrystal (QC) can serve as a natural platform where intrinsic AM is more common and more easily explored.

The QC is a unique long-range-ordered structure without translational symmetry. In QCs, the aperiodic structure and the higher-fold rotational symmetries \cite{levine1984quasi,levine1986quasi,socolar1986quasi} produce various novel phases ~\cite{Wessel2003,Kraus2012,Thiem2015,Andrade2015,Otsuki2016,Koga2017,Huang2018,Huang2019,Longhi2019,Miyazaki2020,caoye2020,Duncan2020,Manna2024,Hua2021,Peng2021,Ghadimi2021,Akihisa2021,Wang2022,Koga2022,Ghosh2023,Chen2023,Yu_Bo_Liu2023,Uri2023,Araujo2024,Yang2024,Yu_Bo_Liu2024,LiuLu2025,chen2025quasi,li2025quasi,AJagannathan1,AJagannathan2,AJagannathan3,Uri2025}, including the magnetic phases~\cite{Wessel2003,Thiem2015,Otsuki2016,Koga2017,Koga2022,Ghosh2023,AJagannathan1,chen2025quasi,li2025quasi}. Now an interesting question arises: Is AM a common magnetic phase in the QC? We shall show that the unique aperiodic structure of QC renders AM more common and more easily explored in QCs than in crystals. Therefore, it is now urgent to classify magnetic phases in QCs based on symmetry to guide the exploration of AM in QCs. Unlike crystals, the QCs lack translational symmetry, making the classification of magnetic phases quite different.

In this paper, we investigate AM driven by interaction in 2D QCs without SOC. We start by showing the difference between the N\'{e}el state on the square lattice and that on a $D_4$-symmetric Thue-Morse QC, with both belonging to the $d_{xy}$-wave IRRP of $D_4$. Consequently, while the former is AFM protected by the combined $\mathcal{P}\mathcal{T}$ and translational symmetry, the lack of translational symmetry in the latter breaks the $\mathcal{P}\mathcal{T}$ symmetry, and the additional mirror or rotation symmetry protects AM. This example demonstrates that AM is more common in QCs than in crystals and is more easily explored through a point-group symmetry-based classification. Therefore, we classify collinear magnetic phases in 2D $D_n$-symmetric QCs, based on 1D IRRPs of $D_n$, including $A_1$, $A_2$, $B_1$, and $B_2$. Symmetry analysis suggests that, while the identity IRRP $A_1$ represents FM and the inversion-odd $B_{1,2}$ IRRPs for twice-of-odd $n$ represent AFM, all the remaining 1D IRRPs represent AM, protected by the mirror-reflection or the rotation symmetry. For verification, we take the Hubbard model as an example and the solved magnetic states in various QCs with different symmetries are all well described by the classification.  Our work sets the QC as a natural platform for intrinsic AM, thereby providing a theoretical foundation for future experimental investigations and potential applications.

\begin{figure}[t]
    \centering
    \includegraphics[width=1\linewidth]{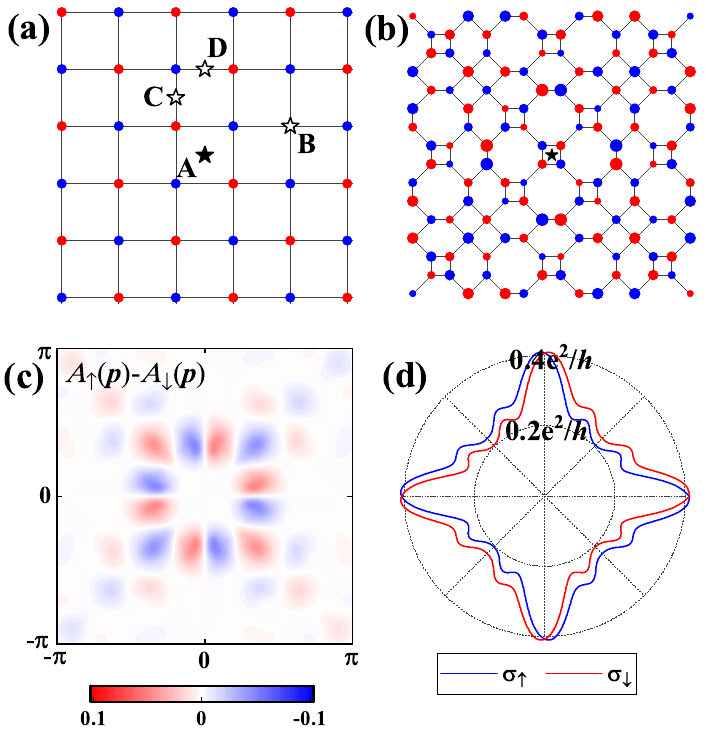}
    \caption{(a-b) N\'eel state in square lattice (a) and $D_4$ symmetric Thue-Morse QC (b). The red (blue) dots represent spin down (up), and the size of each dot represents amplitude of the magnetic moment. Note that in (b), only symmetry related sites host equal amplitude. The solid pentagrams in (a, b) indicate the $D_4$ center, and the hollow pentagrams in (a) indicate three additional inversion centers. (c-d) Two quantities characterizing the AM state on the $D_4$ QC shown in (b). (c) The spin-resolved spectral density difference $\mathcal{A}_{\uparrow}(\bm{p})-\mathcal{A}_{\downarrow}(\bm{p})$ as a function of momentum $\bm{p}$. (d) Spin conductance $\sigma_{\uparrow}$ and $\sigma_{\downarrow}$ as functions of the angle $\phi$ with detecting radius $R=2.5$ in the polar coordinate. The chemical potential is $\mu = -0.5$.} \label{D4_example}
\end{figure}

\paragraph{\textcolor{ZXBlue}{AM in the QC ---}} We start by a comparison between the N\'{e}el states on two $D_4$ symmetric lattices, i.e. the square lattice shown in Fig. \ref{D4_example}(a) and the Thue-Morse QC shown in Fig. \ref{D4_example}(b).  Obviously, the magnetic patterns on both lattices show similar $d_{xy}$-wave (i.e. the $B_2$ IRRP) symmetry behaviors under the $D_4$ point group operations around their symmetry centers marked by the solid pentagrams: both are unchanged under the inversion or mirror reflection about the diagonal directions and change signs under the $c_4^1$ rotation or mirror reflection about the $x$- or $y$- axes $\sigma_{x,y}$. Although both magnetic patterns are inversion-even and hence $\mathcal{P}\mathcal{T}$-odd, the two N\'{e}el states belong to different magnetic classes: In Fig. \ref{D4_example}(a), the $\mathcal{P}\mathcal{T}$ symmetry is recovered after the translation along a unit vector, leading to AFM; while in Fig. \ref{D4_example}(b), the lack of translational symmetry breaks the $\mathcal{P}\mathcal{T}$ symmetry, and the combined $\mathcal{T}$ and $c_4^1$ or $\sigma_{x,y}$ symmetry which guarantees zero net magnetism identifies AM.

The AM in QCs can be characterized by the spin-resolved angle-resolved photoemission spectrum $\mathcal{A}_{\uparrow, \downarrow}(\bm{p})$~\cite{chen2025quasi,li2025quasi,dornellas2025alter} and the spin conductance $\sigma_{\uparrow,\downarrow}(\phi)$ \cite{chen2025quasi}. While $\mathcal{A}_{\uparrow, \downarrow}$ indicates the spectrum weight for spin $\uparrow$ or $\downarrow$ at the Fermi level for the detecting momentum $\bm{p}$, the spin conductance $\sigma_{\uparrow,\downarrow}$ can be measured by placing two tips perpendicular to the QC at $(R\cos\phi,R\sin\phi)$ and $(-R\cos\phi,-R\sin\phi)$, and the measured value is a function of the angle $\phi$, see 
Appendix C 
for details. For the QC N\'{e}el state, the mean-field Hamiltonian is $H_{\text{MF}}=H_{\text{TB}}+\sum_i\Delta_i(n_{i\uparrow}-n_{i\downarrow})$, where the tight-binding (TB) Hamiltonian $H_{\text{TB}}$ is given in 
Appendix A 
and $\{\Delta_{i}\}$ is the magnetic pattern illustrated in Fig. \ref{D4_example}(b).  Solving $H_{\text{MF}}$, the obtained spectrum difference $\mathcal{A}_{\uparrow}-\mathcal{A}_{\downarrow}$ as a function of $\bm{p}$ and the spin conductance difference $\sigma_{\uparrow}-\sigma_{\downarrow}$ as function of $\phi$ in the polar coordinate are shown in Fig.~\ref{D4_example}(c) and (d) respectively, which exhibit nonzero values with $d_{xy}$ symmetry, characteristic of AM. 

The fact that the two N\'{e}el states shown in Fig.~\ref{D4_example}(a) and (b) behave differently under $\mathcal{P}\mathcal{T}$ can be understood from a different but equivalent perspective, clarified here. In Fig.~\ref{D4_example}(a), the square lattice actually hosts three other unequivalent inversion centers marked B, C, and D. One has the flexibility to set B or C as the inversion center, under which the N\'{e}el state is inversion-odd and hence $\mathcal{P}\mathcal{T}$-symmetric. However, in Fig.~\ref{D4_example}(b), the aperiodic structure forbids another inversion center: If a QC hosts two inversion centers, then successive inversion operations around the two centers will lead to a translational symmetry, which is forbidden. Therefore, Fig.~\ref{D4_example}(b) can host only one inversion center, under which the N\'{e}el state is inversion-even and hence breaks the $\mathcal{P}\mathcal{T}$ symmetry.

This example suggests that AM is more common in QCs than in crystals, because QCs cannot provide additional translation operations to recover the broken $\mathcal{P}\mathcal{T}$ symmetry, allowing for the possibility of AM protected by other point-group symmetries. Furthermore, since a QC can host at most one inversion center, it is convenient to explore AM through a point-group symmetry-based classification in the QC, which we shall perform below. 

\begin{table*}[t]
\centering
\caption{Classification of magnetic phases in 2D $D_n$-symmetric QCs without SOC according to the 1D IRRPs of $D_n$. $D_n$ possesses the generating element $c_{n}^{1}$ (rotation) and mirror reflection ($\sigma_x$). The inversion parity $\mathcal{P}$ only applies in cases of even $n$. $l=n/2$ represents the angular momentum. The ``$\surd$'' or ``$\times$'' indicates that the property is satisfied or not. } 
\label{table_irrp}
\begin{tabular}{ccccccccc}
\hline
\hline
$n$ & IRRP & $c_{n}^{1}$ & $\sigma_x$ & $\mathcal{P}$ & Basis functions and symmetry class & Zero magnetism & $\mathcal{P}\mathcal{T}$ & Magnetism \\
\hline
\multirow{4}{*}{$2*$odd}
& $A_1$ & $1$ & $1$ & $1$ & $1$, $s$ & $\times$ & $-1$ & FM \\
& $A_2$ & $1$ & $-1$ & $1$ & $\mathrm{Re}((x+\mathrm{i}y)^{n/2})*\mathrm{Im}((x+\mathrm{i}y)^{n/2})$, e.g. $h*h^{\prime}$ for $n=10$ & $\surd$ & $-1$ & AM \\
& $B_1$ & $-1$ & $1$ & $-1$ & $\mathrm{Re}((x+\mathrm{i}y)^{n/2})$, $l=n/2$, e.g. $h_{x^5-10x^3y^2+5xy^4}$ for $n=10$ & $\surd$ & $+1$ & AFM \\
& $B_2$ & $-1$ & $-1$ & $-1$ & $\mathrm{Im}((x+\mathrm{i}y)^{n/2})$, $l=n/2$, e.g. $h^\prime_{5x^4y-10x^2y^3+y^5}$ for $n=10$ & $\surd$ & $+1$ & AFM \\
\hline
\multirow{4}{*}{$2*$even}
& $A_1$ & $1$ & $1$ & $1$ & $1$, $s$ & $\times$ & $-1$ & FM \\
& $A_2$ & $1$ & $-1$ & $1$ & $\mathrm{Re}((x+\mathrm{i}y)^{n/2})*\mathrm{Im}((x+\mathrm{i}y)^{n/2})$, e.g. $g*g^{\prime}$ for $n=8$ & $\surd$ & $-1$ & AM \\
& $B_1$ & $-1$ & $1$ & $1$ & $\mathrm{Re}((x+\mathrm{i}y)^{n/2})$, $l=n/2$, e.g. $g_{x^4-6x^2y^2+y^4}$ for $n=8$ & $\surd$ & $-1$ & AM \\
& $B_2$ & $-1$ & $-1$ & $1$ & $\mathrm{Im}((x+\mathrm{i}y)^{n/2})$, $l=n/2$, e.g. $g^\prime_{4x^3y-4xy^3}$ for $n=8$ & $\surd$ & $-1$ & AM \\
\hline
\multirow{2}{*}{odd}
& $A_1$ & $1$ & $1$ & $-$ & $1$, $s$ & $\times$ & $-$ & FM \\
& $A_2$ & $1$ & $-1$ & $-$ & $\mathrm{Re}((x+\mathrm{i}y)^n)*\mathrm{Im}((x+\mathrm{i}y)^n)$, e.g. $h*h^{\prime}$ for $n=5$ & $\surd$ & $-$ & AM \\
\hline
\hline
\end{tabular}
\end{table*}

\paragraph{\textcolor{ZXBlue}{Magnetism Classification in QCs ---}}
Magnetic orders in QCs without SOC are classified according to IRRPs of their point groups. The absence of SOC allows for the symmetry operation to act only on the coordinate degree of freedom, without altering the spin. This work specifically focuses on the 1D IRRPs, which describe collinear magnetic orders. Here, we only consider 2D QCs hosting the $D_n$ point group. The magnetism classification according to 1D IRRPs of $D_n$ is listed in Table.~\ref{table_irrp}.

For even $n$, the $D_n$ group possesses four 1D IRRPs: $A_1$, $A_2$, $B_1$ and $B_2$. $A_1$ is the identity IRRP whose basis function is invariant under all symmetry operations of $D_n$; it represents for the $s$-wave magnetic phase. 
Both $B_1$ and $B_2$ IRRPs change sign under a $c_n^1$ rotation, and indicate magnetic phases with the highest angular momentum $l=n/2$. They are distinguished by their behavior under mirror reflection $\sigma_x$ about the $x$-axis: $B_1$ is invariant while $B_2$ reverses sign. Examples of $B_1$ and $B_2$ IRRPs include the two different $g$- ($h$-) wave states in the $D_8$ ($D_{10}$) QC.  
$A_2$ is the product IRRP of $B_1$ and $B_2$; its basis function is invariant under $c_n^1$ but reverses sign under the mirror reflection. Example of $A_2$ includes the $g*g^{\prime}$- ($h*h^{\prime}$-) wave states in the $D_8$ ($D_{10}$) QC.

For odd $n$, the $D_n$ group hosts just two 1D IRRPs: $A_1$ and $A_2$. The corresponding basis functions exhibit similar behaviors under the symmetry operations to those in the case of even $n$. For example, the $A_2$ IRRP of $D_5$ includes the $h*h^{\prime}$-wave magnetic state. 

The properties of the magnetic phases corresponding to each IRRP and their classification is clarified below. 

For the identity IRRP $A_1$ phase, the spin remains invariant under all symmetry operations, resulting in no restriction on the total magnetism. This phase usually hosts a non-zero net magnetism and is FM. 

For phases corresponding to non-identity 1D IRRPs, the spin flips under certain symmetry operations, ensuring zero net magnetism. These phases are classified according to their representations under the $\mathcal{P}\mathcal{T}$ operation. Since a QC can host at most one inversion center, the $\mathcal{P}\mathcal{T}$-even (odd) is equivalent to inversion-odd (even) about this center, convenient for classification.  

From the inversion parity $\mathcal{P}$ listed in Table.~\ref{table_irrp}, it is found that only the $B_1$ and $B_2$ IRRPs for twice-of-odd $n$ are inversion-odd which conserve the $\mathcal{P}\mathcal{T}$ symmetry and thus represent AFM, and all the remaining non-identity 1D IRRPs are inversion-even which break $\mathcal{P}\mathcal{T}$ and represent AM. The AM phases include the $A_2$ IRRPs for all $n$ and the $B_{1,2}$ IRRPs for twice-of-even $n$. In the former case, the AM is protected by mirror-reflection combined with $\mathcal{T}$; in the latter case, both the mirror-reflection and the $c_n^1$ rotation combined with $\mathcal{T}$ can protect AM.



The above results are summarized as: {\color{red}\bf In a 2D $D_n$-symmetric QC, the identity IRRP $A_1$ represents FM; the inversion-odd IRRPs $B_{1,2}$ for twice-of-odd $n$ represent AFM; and all the remaining 1D IRRPs represent AM, protected by mirror reflection or rotation combined with $\mathcal{T}$ symmetries.}  

In the following, we show various exampling states corresponding to the different situations listed in Table \ref{table_irrp}, which are all well described by the classification. 



\paragraph{\textcolor{ZXBlue}{Examples. ---}}
We take the following Hubbard model as an example to investigate magnetic phases in 2D QCs, 
\begin{equation}
\begin{aligned}
H = & H_{\mathrm{TB}} + U \sum_{i} n_{i\uparrow}n_{i\downarrow}
.
\end{aligned}
\end{equation}
Here, $H_{\mathrm{TB}}$ is the TB Hamiltonian given in 
Appendix A, 
$n_{i\sigma}$ is the electron number operator for spin-$\sigma$ and site $i$, 
and $U$ represents the Hubbard repulsion. Open boundary condition is imposed here. To solve this model, we adopt the following linear-response theory detailed in the 
Appendix B, 
which can effectively avoid the problem of being trapped at local energy minima, which is usually encountered in the mean-field study~\cite{chen2025quasi,li2025quasi}.


Considering collinear magnetism with fixed polarization orientation, the real-space distribution of the relative magnetic moment $\{\Delta_{i}\}$ is solved as the eigenvector of the following eigenvalue problem of the spin susceptibility $\chi$ for $U$ approaching its critical value from below,
\begin{equation} \label{eq_chi_eigen}
     \sum_{j} \chi_{ij} \Delta_j =\lambda \Delta_{i}.
\end{equation}
For a given set of Hamiltonian parameters, the realized $\{\Delta_{i}\}$ is that corresponds to the largest eigenvalue $\lambda$. As $\chi$ inherits the point group symmetry of the QC, the eigenvector(s) corresponding to the same eigenvalue form(s) an IRRP of the point group. By tuning the Hamiltonian parameters, we can realize different magnetic phases corresponding to all 1D IRRPs listed in Table \ref{table_irrp}.

\begin{figure}[t]
    \centering
    \includegraphics[width=1\linewidth]{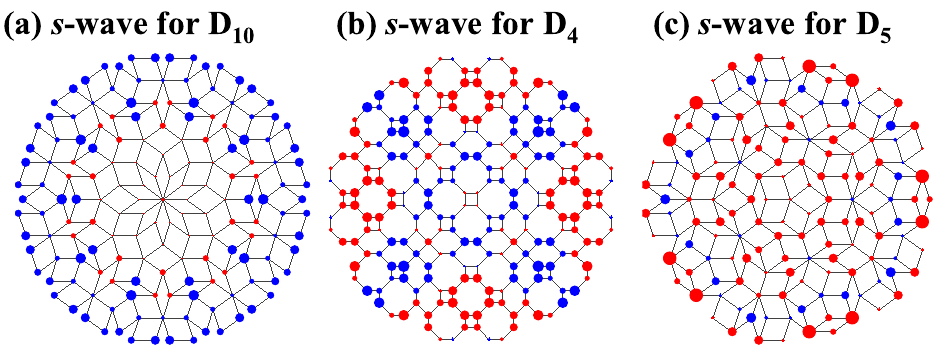}
    \caption{Examples of FM states in decagonal QC (a), Thue-Morse QC (b) and Penrose QC (c). The indications of color and size of the dots are the same as those in Fig.~\ref{D4_example}(b). 
    } \label{FM} 
\end{figure}

To illustrate the classification, we provide examples for each case outlined in Table \ref{table_irrp}. Specifically, $n=10, 6$ are chosen for twice-of-odd $n$, $n=8, 4$ are chosen for twice-of-even $n$, and $n=5$ is chosen for odd $n$. In each case, we tune Hamitonian parameters (see 
Appendix D 
for details) to obtain magnetic states corresponding to all 1D IRRPs through solving Eq. (\ref{eq_chi_eigen}). The magnetic patterns and the corresponding spin-resolved spectrum difference $\mathcal{A}_{\uparrow}(\bm{p})-\mathcal{A}_{\downarrow}(\bm{p})$ and spin conductance $\sigma_{\uparrow}$ and $\sigma_{\downarrow}$ for all these cases are shown in 
Appendix D. 
Consequently, all these examples are well described by the classification. 
Below, we show examples for the FM, AFM and AM, respectively. 

For the FM, Table \ref{table_irrp} suggests that the $s$-wave states corresponding to the identity IRRP $A_1$ for all $n$ are FM. We show some typical $s$-wave states obtained for $n=10,4,5$ corresponding to twice-of-odd, twice-of-even, and odd $n$, respectively. As shown in Fig. \ref{FM}, these states do not change under any symmetry operation in the point group, leading to no restriction to the total magnetism. Therefore, the net magnetism of these states is generally non-zero, leading to FM, verifying Table \ref{table_irrp}. 

\begin{figure}[t]
    \centering
    \includegraphics[width=1\linewidth]{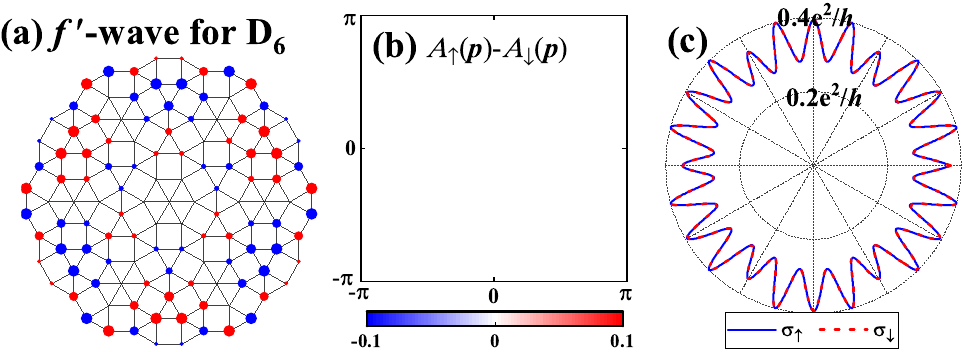}
    \caption{(a) An example of AFM state in a $D_6$ QC. (b)  The spin-resolved spectral density difference $\mathcal{A}_{\uparrow}(\bm{p})-\mathcal{A}_{\downarrow}(\bm{p})$ as a function of momentum $\bm{p}$, showing zero result. (c) Spin conductances $\sigma_{\uparrow}$ and $\sigma_{\downarrow}$ as functions of the angle $\phi$ with $R=3.5$.} \label{AFM}
    
\end{figure}

For the AFM, Table \ref{table_irrp} suggests that the highest-angular-momentum states corresponding to the $B_{1,2}$ IRRPs for twice-of-odd $n$ are AFM. We show a typical $f^{\prime}$-wave (i.e. $B_2$) state for $n=6$. As shown in Fig. \ref{AFM} (a), this state is $\mathcal{P}$ odd and thus $\mathcal{P}\mathcal{T}$ even, which protects the zero net magnetism. The spin-resolved spectrum difference $\mathcal{A}_{\uparrow}(\bm{p})-\mathcal{A}_{\downarrow}(\bm{p})$ (see Fig. \ref{AFM} (b)) and spin conductance difference $\sigma_{\uparrow}-\sigma_{\downarrow}$ (see Fig. \ref{AFM} (c)) for this state are zero, indicating that this state is AFM, verifying Table \ref{table_irrp}.

\begin{figure}[t]
    \centering
    \includegraphics[width=1\linewidth]{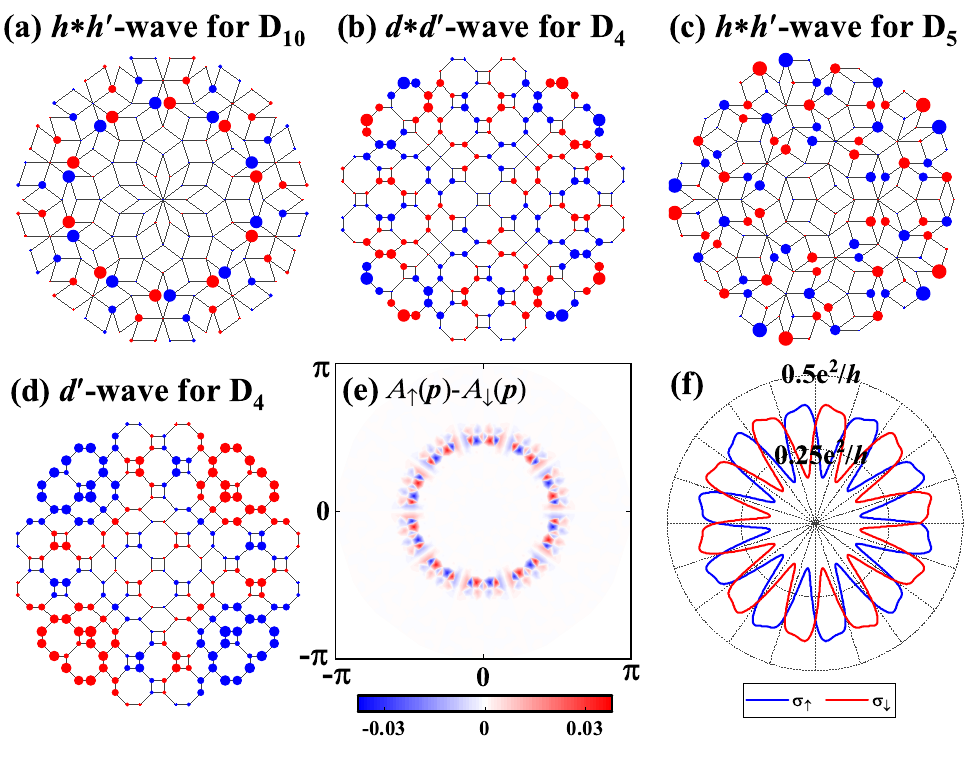}
    \caption{(a-d) Examples of AM states in decagonal QC (a), Thue-Morse QC (b,d) and Penrose QC (c). The indications of color and size of the dots are the same as in Fig.~\ref{D4_example}. (e-f) The $\mathcal{A}_{\uparrow}(\bm{p})-\mathcal{A}_{\downarrow}(\bm{p})$ as a function of momentum $\bm{p}$ (e), and the $\sigma_{\uparrow}$ and $\sigma_{\downarrow}$ as functions of the angle $\phi$ with $R=4.5$ (f), for the AM state in the $D_{10}$ QC shown in (a).} \label{AM}
\end{figure}

For the AM, Table \ref{table_irrp} suggests that the states corresponding to IRRP $A_2$ for all $n$ and IRRPs $B_{1,2}$ for twice-of-even $n$ are AM. For the former case, we show a typical $h*h^{\prime}$-wave state for $n=10$, a typical $d*d^{\prime}$-wave state for $n=4$, and a typical $h*h^{\prime}$-wave state for $n=5$. For the latter case, we show a typical $d^{\prime}$-wave state for $n=4$. As shown in Figs. \ref{AM} (a-d), these states are $\mathcal{P}$ even or do not have $\mathcal{P}$ symmetry, breaking the $\mathcal{P}\mathcal{T}$ symmetry. However, all these states have zero net magnetism protected by mirror reflection. Additionally, the zero net magnetism in the latter case is also protected by $c_4^1$ rotation. The spin resolved spectrum difference $\mathcal{A}_{\uparrow}(\bm{p})-\mathcal{A}_{\downarrow}(\bm{p})$ and spin conductance difference $\sigma_{\uparrow}-\sigma_{\downarrow}$ are non-zero and have the same symmetry as the corresponding state : Figs. \ref{AM} (e-f) show the results for the $h*h^{\prime}$-wave state for $n=10$ as an example, and those for the remaining states are shown in 
Appendix D, 
which are similar, indicating that these states are AM, verifying Table \ref{table_irrp}.

\paragraph{\textcolor{ZXBlue}{Conclusion and Discussion. ---}} 
We have shown that AM is more common in QCs than in crystals due to lack of translational symmetry, and have further classified magnetic phases in 2D $D_n$-symmetric QCs according to the 1D IRRPs of $D_n$. Our results suggest that all 1D non-identity IRRPs are AM, except the inversion-odd $B_{1,2}$ IRRPs for twice-of-odd $n$. This result is widely verified in the exampling study of the Hubbard model in various QCs with different symmetries.


This work primarily focuses on the collinear magnetic states governed by non-trivial 1D IRRPs of the $D_n$ group in 2D QCs.
These findings can be naturally extended to 3D QC systems, e.g. with $I_h$ point group.
Additionally, other collinear magnetic states transforming under higher-dimensional IRRPs may also be AM. 
Furthermore, higher-dimensional IRRPs in 2- or 3- spatial dimensions can lead to unconventional magnetic phases such as nematic or chiral phases. A full investigation of these states will be the subject of future research.

~~~~~~~~~~~~~~~~

\paragraph{\textcolor{ZXBlue}{Acknowledgement. ---}} 
We are grateful to the stimulating discussions with Lunhui Hu, Cheng-Cheng Liu and Wanxiang Feng. F.Y. is supported by the national natural science foundation of China under the Grant Nos. 12234016, 12074031.
Z.P. is supported by the startup funding from Xiamen University.  C.L. is
supported by the National Natural Science Foundation
of China under the Grants No. 12304180.


~~~~
\appendix

\onecolumngrid
\renewcommand{\theequation}{S\arabic{equation}}
\renewcommand{\thefigure}{S\arabic{figure}}
\setcounter{equation}{0}
\setcounter{figure}{0}


\section{The Tight-Binding Hamiltonian}

\begin{figure}[htbp]
    \centering
    \includegraphics[width=1\linewidth]{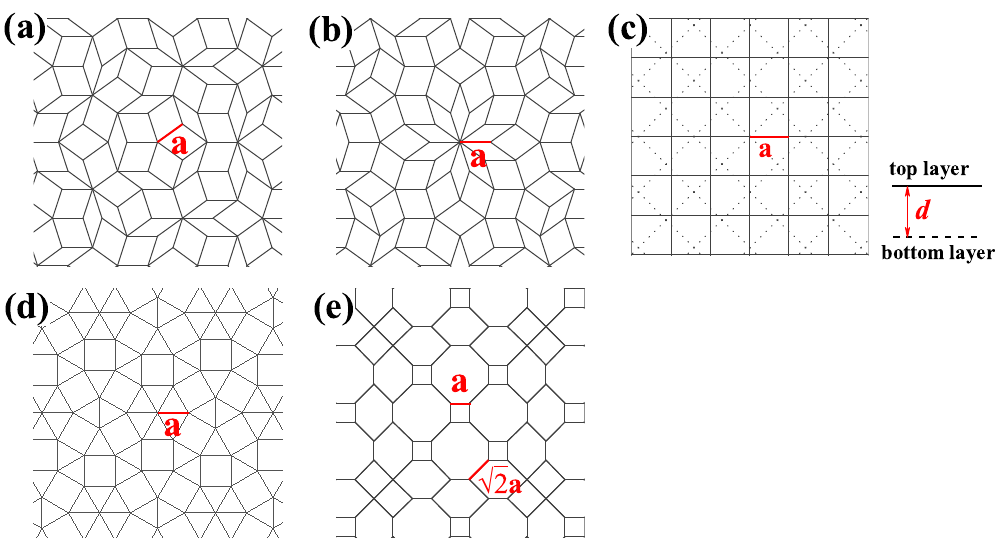}
    \caption{The five QC structures we studied, where the red line marks the length unit $a =1$. (a) Penrose QC with $D_5$ symmetry. (b) decagonal QC with $D_{10}$ symmetry. (c) $45^{\circ}$-twisted bilayer square lattice with $D_8$ symmetry. The solid(dashed) line represents the top(bottom) layer. The interlayer distance $d$ is marked at the bottom right corner. (d) square-triangle tiling QC with $D_6$ symmetry (e) Thue-Morse QC with $D_4$ symmetry. } \label{QC}
\end{figure}

As the beginning of our study, we construct the tight-binding (TB) model in the 2D quasicrystals (QCs). The TB Hamiltonian is given by:

\begin{equation}
H_{\text{TB}}=-\sum_{i\neq j,\sigma}\mathrm{e}^{-r_{ij}/a}c^{\dagger}_{i\sigma}c_{j\sigma}-\mu\sum_{i,\sigma}n_{i\sigma}.
\end{equation}
Here $c^{(\dagger)}_{i\sigma}$ annihilates (creates) a spin-$\sigma$ electron at site $i$, $r_{ij}$ is the distance between site $i$ and site $j$, $a$ is the length unit (marked in Fig. \ref{QC}), 
$\mu$ is the chemical potential,
$n_{i\sigma}$ denotes the number operator for spin-$\sigma$ electrons at site $i$. 

The five QC structures we studied are shown in Fig. \ref{QC}, including the Penrose QC ($D_5$ point group, shown in (a) panel), decagonal QC ($D_{10}$ point group, shown in (b) panel), $45^{\circ}$-twisted bilayer square lattice ($D_{8}$ point group, shown in (c) panel), square-triganle tiling QC ($D_{6}$ point group, shown in (d) panel), and Thue-Morse QC ($D_{4}$ point group, shown in (e) panel). The length unit $a$ for each QC is marked on the corresponding panel. Specifically, for the $45^{\circ}$-twisted bilayer square lattice, there is also an interlayer distance $d$ shown in Fig. \ref{QC} (c), which is set to $d$ = $2a$ in our calculations.

\section{Linear Response Theory for the Magnetic Order}

The magnetic order can be effectively obtained using linear response theory. 
We begin with a Hamiltonian perturbed by a small external magnetic field $\bm{h}_{i}$ along the $z$-direction,
\begin{equation}
    H_{h} = H_{0} + \sum_{i} \bm{h}_{i} \cdot \bm{S}_{i}.
\end{equation}
Here, $H_{0}$ is the unperturbed Hamiltonian, 
$\bm{S}_{i}$ is the spin at site $i$, 
and $\bm{h}_{i}\rightarrow0$ represents the external magnetic field. 
According to perturbation theory, the perturbed ground state $\left|\psi_{G}\right\rangle$ of $H_h$ up to the first-order is expressed as
\begin{equation}
    \left|\psi_{G}\right\rangle = \left|\psi_{G}^{(0)}\right\rangle + \sum_{n \neq G} \frac{\left\langle\psi_{n}^{(0)}\right|H_h\left|\psi_{G}^{(0)}\right\rangle}{E_{G}^{(0)}-E_{n}^{(0)}} \left|\psi_{n}^{(0)}\right\rangle, 
\end{equation}
where $\left|\psi_{G}^{(0)}\right\rangle$ and $E_{G}^{(0)}$ are the unperturbed ground state and energy of $H_{0}$, 
and $\left|\psi_{n}^{(0)}\right\rangle$ and $E_{n}^{(0)}$ are the $n$-th excited eigenstates and eigenvalues of $H_{0}$. 
To determine the magnetic order, we evaluate the ground state expectation value of the spin operator at site $i$,
\begin{equation}
    \begin{aligned}
        \left\langle\bm{S}_{i}\right\rangle = \left\langle\psi_{G}\right|\bm{S}_{i}\left|\psi_{G}\right\rangle 
        = \left\langle\psi_{G}^{(0)}\right|\bm{S}_{i}\left|\psi_{G}^{(0)}\right\rangle + \sum_{j, n \neq G} \left( \frac{\left\langle\psi_{G}^{(0)}\right|\bm{S}_{i}\left|\psi_{n}^{(0)}\right\rangle\left\langle\psi_{n}^{(0)}\right|\bm{h}_{j}\cdot\bm{S}_{j}\left|\psi_{G}^{(0)}\right\rangle}{E_{G}^{(0)}-E_{n}^{(0)}} + \mathrm{c.c.} \right) + O(h^2). 
    \end{aligned}
\end{equation}
Here, $O(h^2)$ is higher-order contribution and can be ignored in the limit $\bm{h}_i\rightarrow 0$. 
The first term, $\left\langle\psi_{G}^{(0)}\right|\bm{S}_{i}\left|\psi_{G}^{(0)}\right\rangle=0$, vanishes since the unperturbed ground state of $H_{0}$ is spin $\mathrm{SU(2)}$ symmetric,
while the spin operator $\bm{S}_{i}$ breaks this symmetry. 
Therefore, the leading order contribution is given by 
\begin{equation}
    \left\langle\bm{S}_{i}\right\rangle = \sum_{j, n \neq G} \left( \frac{\left\langle\psi_{G}^{(0)}\right|\bm{S}_{i}\left|\psi_{n}^{(0)}\right\rangle\left\langle\psi_{n}^{(0)}\right|\bm{h}_{j}\cdot\bm{S}_{j}\left|\psi_{G}^{(0)}\right\rangle}{E_{G}^{(0)}-E_{n}^{(0)}} + \mathrm{c.c.} \right). 
\end{equation}
By applying linear response theory, we can relate the induced spin polarization to the external field, $\left\langle\bm{S}_i\right\rangle\propto\bm{h}_i$. 
Specifically, for a magnetic field $\bm{h}_i$ pointing along the $z$-direction, the induced spin polarization along $z$-axis, $S^z$, is proportional to the field strength, $h_i^z$.
This relationship leads to the following expression,
\begin{equation} \label{eq_Si}
    \begin{aligned}
        \left\langle{}S_{i}^z\right\rangle & = \sum_{j, n \neq G} \left( \frac{\left\langle\psi_{G}^{(0)}\right|S_{i}^z\left|\psi_{n}^{(0)}\right\rangle\left\langle\psi_{n}^{(0)}\right|h_{j}^z{}S_{j}^z\left|\psi_{G}^{(0)}\right\rangle}{E_{G}^{(0)}-E_{n}^{(0)}} + \mathrm{c.c.} \right) \\
        & = \sum_{j} 2 \alpha * \mathrm{Re} \left( \sum_{n \neq G} \frac{\left\langle\psi_{G}^{(0)}\right|S_{i}^z\left|\psi_{n}^{(0)}\right\rangle\left\langle\psi_{n}^{(0)}\right|S_{j}^z\left|\psi_{G}^{(0)}\right\rangle}{E_{G}^{(0)}-E_{n}^{(0)}} \right) \left\langle{}S_{j}^{z}\right\rangle. 
    \end{aligned}
\end{equation}
From Eq. (\ref{eq_Si}), the magnetic order, represented by the vector of spin expectation values, corresponds to an eigenvector of the matrix, defined by the term within the parenthesis. 

To analyze the magnetic order, we first introduce the Matsubara Green's function defined by,
\begin{equation}
    \mathcal{G}_{ij}(\tau)=\left\langle{}T_{\tau}S_{i}^{z}(\tau)S_{j}^{z}(0)\right\rangle, 
\end{equation}
where $\tau$ is the imaginary time and $T_{\tau}$ represents the time-ordered product of operators. 
Its Fourier transform gives the static spin susceptibility $\chi_{ij}$,
\begin{equation}
    \chi_{ij}\equiv\mathcal{G}_{ij}(\mathrm{i}\omega_n=0),\qquad
    \mathcal{G}_{ij}(\mathrm{i}\omega_n) = \int_{0}^{\beta} \mathrm{d}\tau \mathcal{G}_{ij}(\tau)\mathrm{e}^{\mathrm{i}\omega_n\tau}, 
\end{equation}
where $\beta$ is the inverse temperature and $\omega_n=\frac{2\pi{}n}{\beta}$ is the Matsubara frequency. 
The Lehmann's representation of $\chi_{ij}$ is precisely the summation in the braket in Eq. (\ref{eq_Si}). Therefore, if $\chi_{ij}$ is seen as a matrix element where $i$ and $j$ are the row and column indices, respectively, the magnetic order is obtained from the eigenvector of the real part of $\chi$ matrix. 

The spin susceptibility can be obtained as follows. First we obtain the bare susceptibility $\chi^{(0)}$, whose definition is similar to the above but the expectation is obtained over the bare TB Hamiltonian. 
Then the expression of $\chi^{(0)}$ is 
\begin{equation}
    \chi^{(0)}_{ij} = \sum_{mn} \xi_{im} \xi_{in} \xi_{jm} \xi_{jn} \frac{n_F(\varepsilon_m-\mu)-n_F(\varepsilon_n-\mu)}{\varepsilon_n-\varepsilon_m}, 
\end{equation}
where $n_F$ represents the Fermi-Dirac distribution function, $\varepsilon_m$ is the $m$-th eigenvalue of $H_{\mathrm{TB}}$, $\xi_{im}$ is the $i$-th value in the eigenvector of $H_{\mathrm{TB}}$ belonging to $\varepsilon_m$, and $\mu$ is the chemical potential. 
Then the renormalized spin susceptibility for the interacting case up to the random-phase approximation level is obtained from the Dyson's equation, 
\begin{equation} \label{eq_chi}
    \chi = \left(I-U\chi^{(0)}\right)^{-1}\chi^{(0)}. 
\end{equation}
When the interaction strength is larger than the critical value $U_c$, $\chi$ diverges, forming a magnetic order. The magnetic pattern is obtained from the eigenvector belonging to the largest eigenvalue of $\chi$ for $U\rightarrow{}U_c^{-}$. 
For this single-orbital case, $U$ in the Dyson's equation is just a number and equals the strength of the onsite intraorbital Coulomb repulsion. Thus, $\chi$ commutes with $\chi^{(0)}$. 

\section{Characterization of Altermagnetism}
The altermagnetism (AM) in QCs can be characterized by the spin-resolved angle-resolved photoemission spectrum \cite{chen2025quasi,li2025quasi,dornellas2025alter} and the spin conductance \cite{chen2025quasi}. 

To observe these quantities, we start with a mean-field Hamiltonian
\begin{equation}
    H_{\mathrm{MF}} = H_{\mathrm{TB}} + \sum_{i} \Delta_{i} \left(n_{i\uparrow}-n_{i\downarrow}\right), 
\end{equation}
where 
$\left\{\Delta_{i}\right\}$ is the magnetic pattern obtained by solving Eq. (2) in the main text. We note that 
only the region near the rotational center are shown in all figures in the main text and here.
In the Wannier basis, the matrix form of $H_{\mathrm{MF}}$ is a block diagonal matrix, 
\begin{equation}
    (H_{\mathrm{MF}}) = \left(
    \begin{array}{c|c}
        (H_{\uparrow}) & 0 \\
        \hline
        0 & (H_{\downarrow})
    \end{array}
    \right). 
\end{equation}
Here, $(H_{\mathrm{MF}})$ is the matrix form of $H_{\mathrm{MF}}$ in the Wannier basis, and $(H_{\sigma})$ is the matrix form of the mean-field Hamiltonian for spin $\sigma$ ($\sigma=\uparrow,\downarrow$), which is defined as
\begin{equation}
    H_{\sigma} = -\sum_{i\neq j}\mathrm{e}^{-r_{ij}/a}c^{\dagger}_{i\sigma}c_{j\sigma}-\mu\sum_{i}n_{i\sigma} + \sigma \sum_{i} \Delta_{i} n_{i\sigma}. 
\end{equation}
In the last term, $\sigma=+1$ for spin up and $\sigma=-1$ for spin down. 
Then we solve the eigenvalues and eigenvectors of $(H_{\sigma})$
\begin{equation}
    \sum_{j} (H_{\sigma})_{ij} \xi_{jm}^{\sigma} = E_{m}^{\sigma} \xi_{im}^{\sigma}, 
\end{equation}
where $(H_{\sigma})_{ij}$ is the matrix element at the $i$-th row and the $j$-th column in $(H_{\sigma})$, $E_m^{\sigma}$ is the $m$-th eigenvalue of $(H_{\sigma})$, and $\xi_{im}^{\sigma}$ is the $i$-th value in the $m$-th eigenvector of $(H_{\sigma})$. 

We consider the spectrum weight for spin $\sigma$ at the Fermi level, which is expressed as
\begin{equation}
    \mathcal{A}_{\sigma}(\bm{p}) = -\frac{\eta}{\pi} \frac{\left|\left\langle{}\bm{p}\sigma|m\sigma\right\rangle\right|^2}{\left(E_F-E_m^{\sigma}\right)^2+\eta^2}, 
\end{equation}
where $E_F$ is the Fermi level, $\left|m\sigma\right\rangle$ is the $m$-th eigenstate of $H_{\sigma}$, $\left|\bm{p}\sigma\right\rangle=\frac{1}{\sqrt{N}}\sum_{i}\mathrm{e}^{\mathrm{i}\bm{p}\cdot\bm{r}_{i}}c^{\dagger}_{i\sigma}\left|0\right\rangle$ ($N$ is site number, $\bm{r}_i$ is the position of site $i$ and $\left|0\right\rangle$ is vaccum state) is the plane wave with momentum $\bm{p}$, and $\eta$ is a small value. 

The spin conductance is calculated by the Landauer-B\"uttiker formula \cite{landauer1970,buttiker1988,fisher1981} and the recursive Green's function method \cite{mackinnon1985calculation,metalidis2005}, see Ref. \cite{chen2025quasi}. 

\section{More Results}

Here we show the results corresponding to different cases in Table I in the main text. We consider $n=10,6$ for twice-of-odd $n$, $n=8,4$ for twice-of-even $n$, and $n=5$ for odd $n$ as examples. All 1D irreducible representations (IRRPs), $A_1$, $A_2$, $B_1$ and $B_2$ for even $n$ or $A_1$ and $A_2$ for odd $n$, are considered. All magnetic states we will present are calculated based on Eq. (2) in the main text.

\begin{figure}[htbp]
    \centering
    \includegraphics[width=0.8\linewidth]{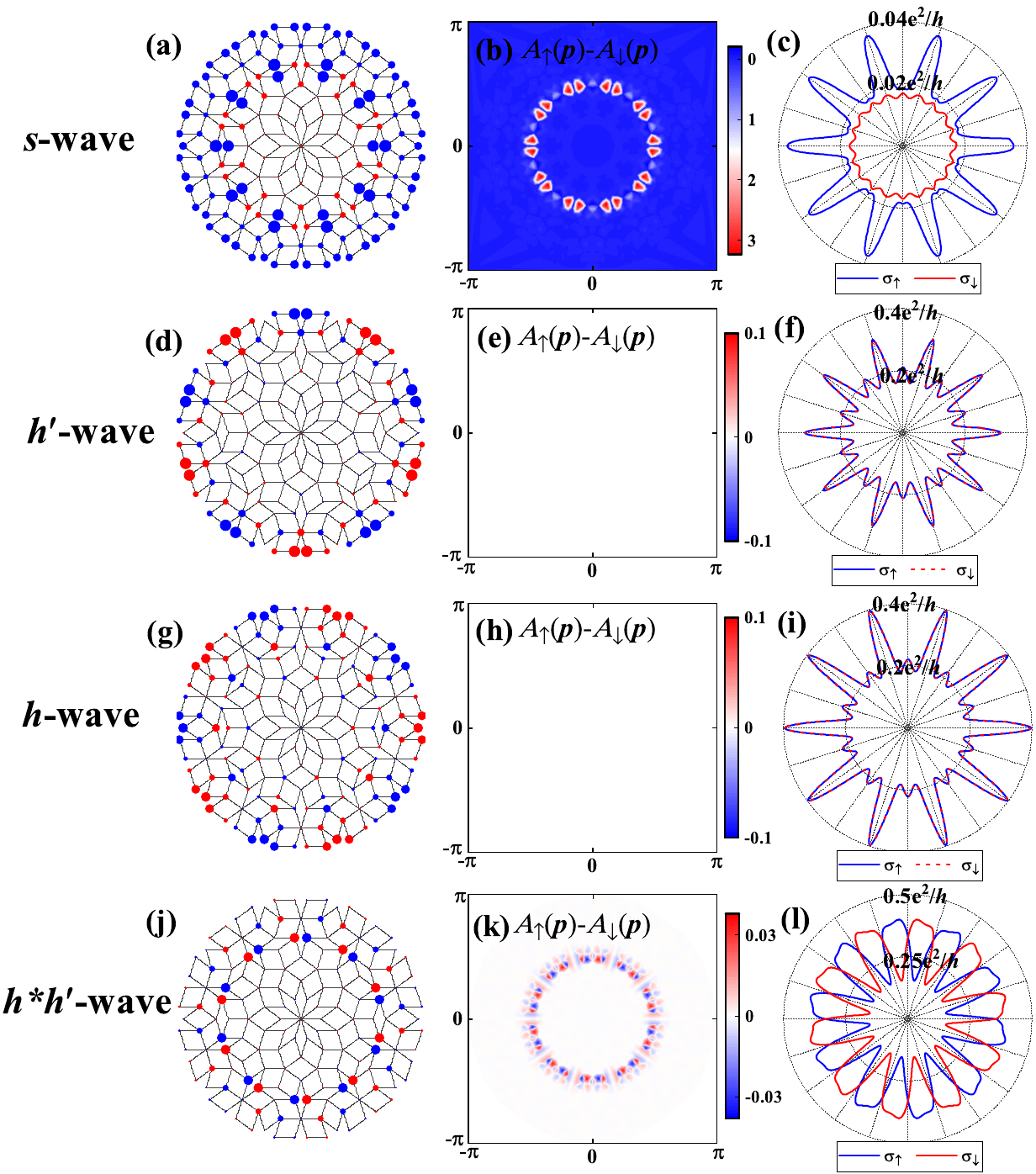}
    \caption{Magnetic states in the decagonal QC with $D_{10}$ symmetry and their properties. (a) The spatial distribution of local magnetic order $\Delta_i$ with $s$-wave symmetry in real space. The color and area of the dots represent the sign and relative magnitude of the magnetic order, respectively. (b) The spin-resolved spectral density $\mathcal{A}_{\uparrow}(\bm{p})-\mathcal{A}_{\downarrow}(\bm{p})$ as a function of momentum $\bm{p}$. (c) Spin conductances $\sigma_{\uparrow}$ and $\sigma_{\downarrow}$ as functions of the angle $\phi$ with $R=4.5$. The value of $\sigma$ is represented by the radial distance. (d-f), (g-i) and (j-l) are similar to (a-c), except that the symmetry of the shown state is $h^{\prime}$-, $h$-, and $h*h^{\prime}$-wave, respectively. The chemical potential is fixed at $\mu = -0.487$ in (a-c) and (j-l) panels, and $\mu = -0.325$ in (d-i) panels.}
    \label{fig_SD10}
\end{figure}

The results for $n=10$ are shown in Fig. \ref{fig_SD10}. 
The magnetic pattern for a typical $s$-wave state, corresponding to the identity IRRP $A_1$, and the spin-resolved spectrum difference $\mathcal{A}_{\uparrow}(\bm{p})-\mathcal{A}_{\downarrow}(\bm{p})$ and spin conductance $\sigma_{\uparrow}$ and $\sigma_{\downarrow}$ for this state are shown in Figs. \ref{fig_SD10} (a-c), respectively. As shown in Fig. \ref{fig_SD10} (a), this state does not change under any symmetry operation in the point group, leading to no restriction to the total magnetism, thus this state is ferromagnetism (FM). The spin-resolved spectrum difference $\mathcal{A}_{\uparrow}(\bm{p})-\mathcal{A}_{\downarrow}(\bm{p})$ (Fig. \ref{fig_SD10} (b)) and spin conductance difference $\sigma_{\uparrow}-\sigma_{\downarrow}$ are non-zero and also do not change under any symmetry operation, also indicating that the state is FM. These results verify Table I in the main text. 

The magnetic patterns for a typical $h^{\prime}$-wave state and a typical $h$-wave state, corresponding to IRRP $B_2$ and $B_1$, respectively, and the spin-resolved spectrum difference $\mathcal{A}_{\uparrow}(\bm{p})-\mathcal{A}_{\downarrow}(\bm{p})$ and spin conductance $\sigma_{\uparrow}$ and $\sigma_{\downarrow}$ for these states are shown in Figs. \ref{fig_SD10} (d-i). As shown in Figs. \ref{fig_SD10} (d) and (g), both of these states are parity ($\mathcal{P}$) odd and thus $\mathcal{P}$-time ($\mathcal{T}$) even, protecting the zero total magnetism. The spin-resolved spectrum difference $\mathcal{A}_{\uparrow}(\bm{p})-\mathcal{A}_{\downarrow}(\bm{p})$ (Figs. \ref{fig_SD10} (e) and (h)) and spin conductance difference $\sigma_{\uparrow}-\sigma_{\downarrow}$ (Figs. \ref{fig_SD10} (f) and (i)) for these states are all zero, indicating that these states are both antiferromagnetism (AFM). These results verify Table I in the main text. 

The magnetic pattern for a typical $h*h^{\prime}$-wave state, corresponding to IRRP $A_2$, and the spin-resolved spectrum difference $\mathcal{A}_{\uparrow}(\bm{p})-\mathcal{A}_{\downarrow}(\bm{p})$ and spin conductance $\sigma_{\uparrow}$ and $\sigma_{\downarrow}$ for this state are shown in Figs. \ref{fig_SD10} (j-l), respectively. As shown in Fig. \ref{fig_SD10} (j), this state is $\mathcal{P}$ even and thus $\mathcal{P}\mathcal{T}$ odd. However, this state has zero total magnetism protected by mirror reflection. The spin-resolved spectrum difference $\mathcal{A}_{\uparrow}(\bm{p})-\mathcal{A}_{\downarrow}(\bm{p})$ (Fig. \ref{fig_SD10} (k)) and spin conductance difference $\sigma_{\uparrow}-\sigma_{\downarrow}$ (Fig. \ref{fig_SD10} (l)) are non-zero and have the same symmetry as the corresponding state, indicating that this state is AM. These results verify Table I in the main text. 


\begin{figure}[htbp]
    \centering
    \includegraphics[width=0.8\linewidth]{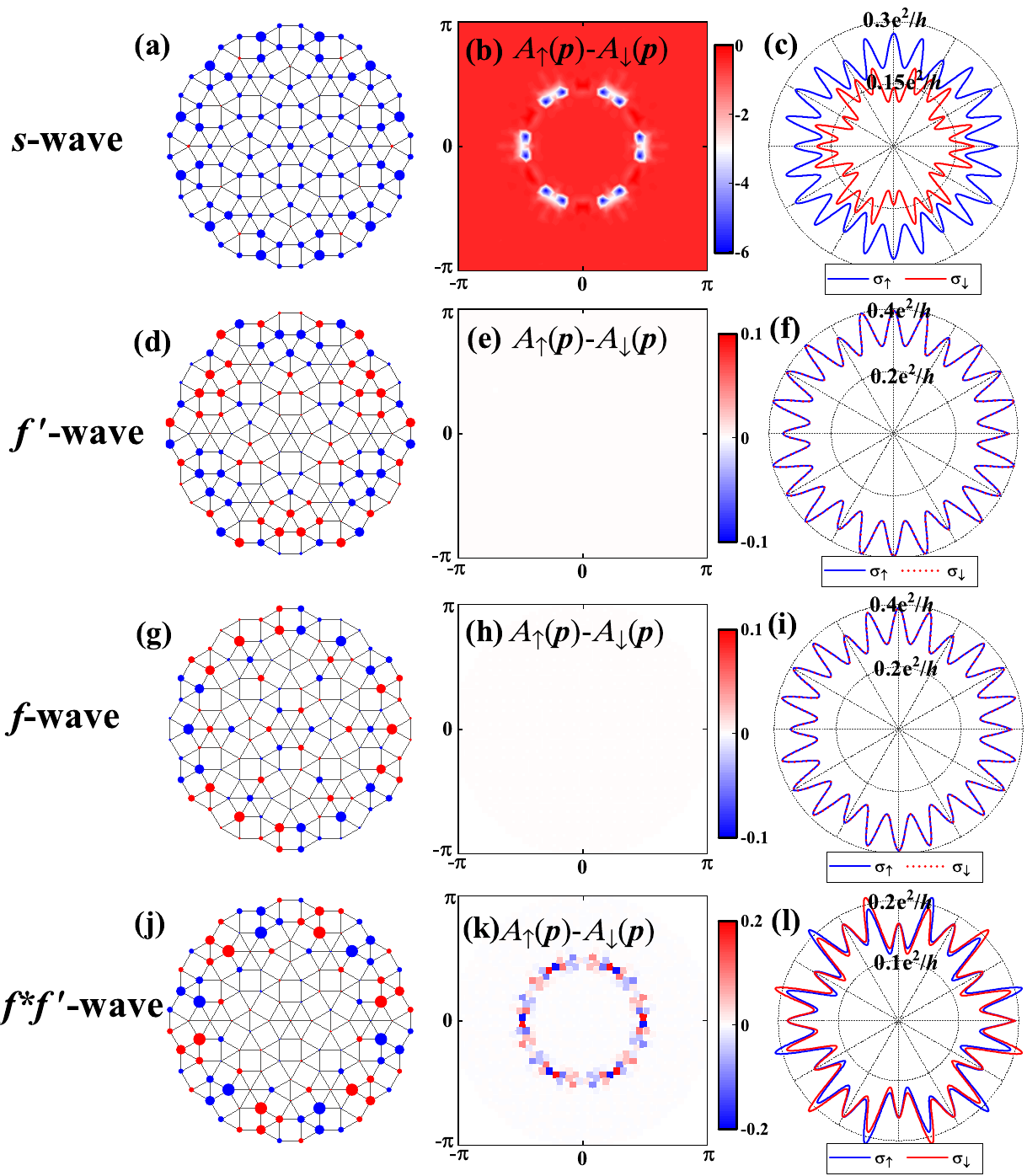}
    \caption{Magnetic states in the square-triangle tiling QC with $D_6$ symmetry and their properties. (a) The spatial distribution of local magnetic order $\Delta_i$ with $s$-wave symmetry in real space. The color and area of the dots represent the sign and relative magnitude of the magnetic order, respectively. (b) The spin-resolved spectral density $\mathcal{A}_{\uparrow}(\bm{p})-\mathcal{A}_{\downarrow}(\bm{p})$ as a function of momentum $\bm{p}$. (c) Spin conductances $\sigma_{\uparrow}$ and $\sigma_{\downarrow}$ as functions of the angle $\phi$ with $R=3.5$. The value of $\sigma$ is represented by the radial distance. (d-f), (g-i) and (j-l) are similar to (a-c), except that the symmetry of the shown state is $f^{\prime}$-, $f$-, and $f*f^{\prime}$-wave, respectively. The chemical potential is fixed at $\mu = -0.365$ in (a-c) and (j-l) panels, and $\mu = -0.246$ in (d-i) panels.} \label{fig_SD6}
\end{figure}

The results for $n=6$ are shown in Fig. \ref{fig_SD6}. 
The magnetic pattern for a typical $s$-wave state, corresponding to the identity IRRP $A_1$, and the spin-resolved spectrum difference $\mathcal{A}_{\uparrow}(\bm{p})-\mathcal{A}_{\downarrow}(\bm{p})$ and spin conductance $\sigma_{\uparrow}$ and $\sigma_{\downarrow}$ for this state are shown in Figs. \ref{fig_SD6} (a-c), respectively. As shown in Fig. \ref{fig_SD6} (a), this state does not change under any symmetry operation in the point group, leading to no restriction to the total magnetism, thus this state is FM. The spin-resolved spectrum difference $\mathcal{A}_{\uparrow}(\bm{p})-\mathcal{A}_{\downarrow}(\bm{p})$ (Fig. \ref{fig_SD6} (b)) and spin conductance difference $\sigma_{\uparrow}-\sigma_{\downarrow}$ are non-zero and also do not change under any symmetry operation, also indicating that the state is FM. These results verify Table I in the main text. 

The magnetic patterns for a typical $f^{\prime}$-wave state and a typical $f$-wave state, corresponding to IRRP $B_2$ and $B_1$, respectively, and the spin-resolved spectrum difference $\mathcal{A}_{\uparrow}(\bm{p})-\mathcal{A}_{\downarrow}(\bm{p})$ and spin conductance $\sigma_{\uparrow}$ and $\sigma_{\downarrow}$ for these states are shown in Figs. \ref{fig_SD6} (d-i). As shown in Figs. \ref{fig_SD6} (d) and (g), both of these states are $\mathcal{P}$ odd and thus $\mathcal{P}\mathcal{T}$ even, protecting the zero total magnetism. The spin-resolved spectrum difference $\mathcal{A}_{\uparrow}(\bm{p})-\mathcal{A}_{\downarrow}(\bm{p})$ (Figs. \ref{fig_SD6} (e) and (h)) and spin conductance difference $\sigma_{\uparrow}-\sigma_{\downarrow}$ (Figs. \ref{fig_SD6} (f) and (i)) for these states are all zero, indicating that these states are both AFM. These results verify Table I in the main text. 

The magnetic pattern for a typical $f*f^{\prime}$-wave state, corresponding to IRRP $A_2$, and the spin-resolved spectrum difference $\mathcal{A}_{\uparrow}(\bm{p})-\mathcal{A}_{\downarrow}(\bm{p})$ and spin conductance $\sigma_{\uparrow}$ and $\sigma_{\downarrow}$ for this state are shown in Figs. \ref{fig_SD6} (j-l), respectively. As shown in Fig. \ref{fig_SD6} (j), this state is $\mathcal{P}$ even and thus $\mathcal{P}\mathcal{T}$ odd. However, this state has zero total magnetism protected by mirror reflection. The spin-resolved spectrum difference $\mathcal{A}_{\uparrow}(\bm{p})-\mathcal{A}_{\downarrow}(\bm{p})$ (Fig. \ref{fig_SD6} (k)) and spin conductance difference $\sigma_{\uparrow}-\sigma_{\downarrow}$ (Fig. \ref{fig_SD6} (l)) are non-zero and have the same symmetry as the corresponding state, indicating that this state is AM. These results verify Table I in the main text. 

\begin{figure}[htbp]
    \centering
    \includegraphics[width=0.8\linewidth]{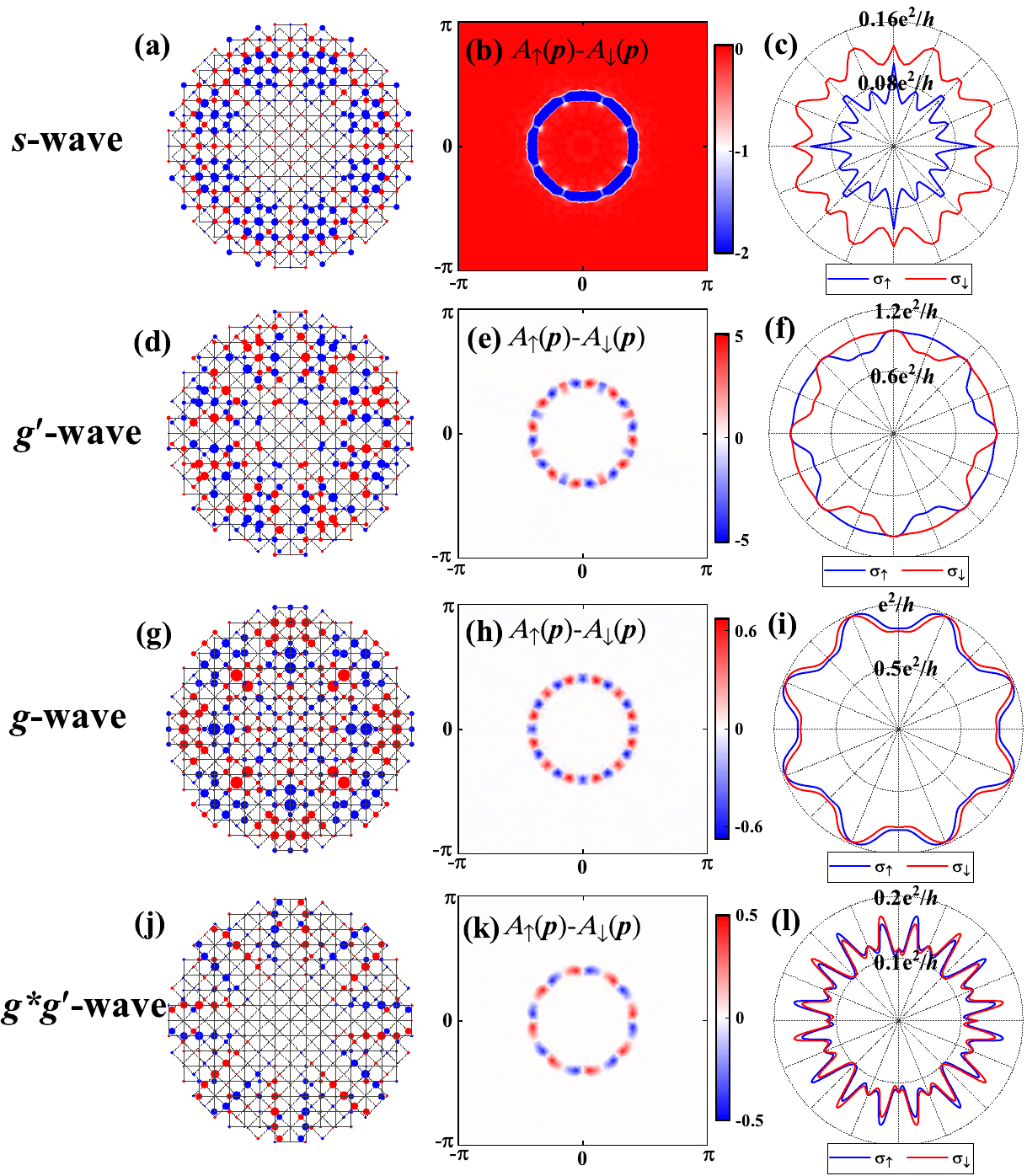}
    \caption{Magnetic states in the $45^{\circ}$-twisted bilayer square lattice with $D_8$ symmetry and their properties. (a) The spatial distribution of local magnetic order $\Delta_i$ with $s$-wave symmetry in real space. The color and area of the dots represent the sign and relative magnitude of the magnetic order, respectively. (b) The spin-resolved spectral density $\mathcal{A}_{\uparrow}(\bm{p})-\mathcal{A}_{\downarrow}(\bm{p})$ as a function of momentum $\bm{p}$. (c) Spin conductances $\sigma_{\uparrow}$ and $\sigma_{\downarrow}$ as functions of the angle $\phi$ with $R=2.5$. The value of $\sigma$ is represented by the radial distance. (d-f), (g-i) and (j-l) are similar to (a-c), except that the symmetry of the shown state is $g^{\prime}$-, $g$-, and $g*g^{\prime}$-wave, respectively. The chemical potential is fixed at $\mu = -0.942$ in all panels.} \label{fig_SD8}
\end{figure}

For twice-of-even $n$, we show some results of the magnetic states in the $45^{\circ}$-twisted bilayer square lattice with $D_8$ symmetry and their properties in Fig. \ref{fig_SD8}. $D_8$ group has four 1D IRRPs, $A_1$, $A_2$, $B_1$ and $B_2$, corresponding to $s$-wave, $g*g^{\prime}$-wave, $g$-wave, and $g^{\prime}$-wave phases, respectively. The $s$-wave phase has non-zero total magnetism and thus it is a FM state.
The other phases have zero total magnetism and no $\mathcal{P}\mathcal{T}$ symmetry, which are the features of AM state.

The magnetic pattern for a typical $s$-wave state, corresponding to the identity IRRP $A_1$, and the spin-resolved spectrum difference $\mathcal{A}_{\uparrow}(\bm{p})-\mathcal{A}_{\downarrow}(\bm{p})$ and spin conductance $\sigma_{\uparrow}$ and $\sigma_{\downarrow}$ for this state are shown in Figs. \ref{fig_SD8} (a-c), respectively. As shown in Fig. \ref{fig_SD8} (a), this state does not change under any symmetry operation in the $D_8$ point group, leading to no restriction to the total magnetism, thus it is a FM state. The spin-resolved spectrum difference $\mathcal{A}_{\uparrow}(\bm{p})-\mathcal{A}_{\downarrow}(\bm{p})$ (Fig. \ref{fig_SD8} (b)) and spin conductance difference $\sigma_{\uparrow}-\sigma_{\downarrow}$ (Fig. \ref{fig_SD8} (c)) are non-zero and also do not change under any symmetry operation. These results verify Table I in the main text. 


The patterns of a typical $g^{\prime}$-wave, $g$-wave and $g*g^{\prime}$-wave states are shown in Figs.~\ref{fig_SD8} (d,g,j), and their properties are shown in Figs. \ref{fig_SD8} (e-f), (h-i) and (k-l), respectively. As shown in Figs. \ref{fig_SD8} (d), (g) and (j), all of these states are $\mathcal{P}$ even and thus $\mathcal{PT}$ odd but have zero total magnetism protected by mirror reflection or $c_8^1$ rotation. The spin-resolved spectrum $\mathcal{A}_{\uparrow}-\mathcal{A}_{\downarrow}$ (Figs. \ref{fig_SD8} (e), (h) and (k)) and spin conductance difference $\sigma_{\uparrow}-\sigma_{\downarrow}$ (Figs. \ref{fig_SD8} (f), (i) and (l)) are non-zero and have the same symmetry as the corresponding state, indicating that all these states are AM. All the results verify Table I in the main text.


\begin{figure}[htbp]
    \centering
    \includegraphics[width=0.8\linewidth]{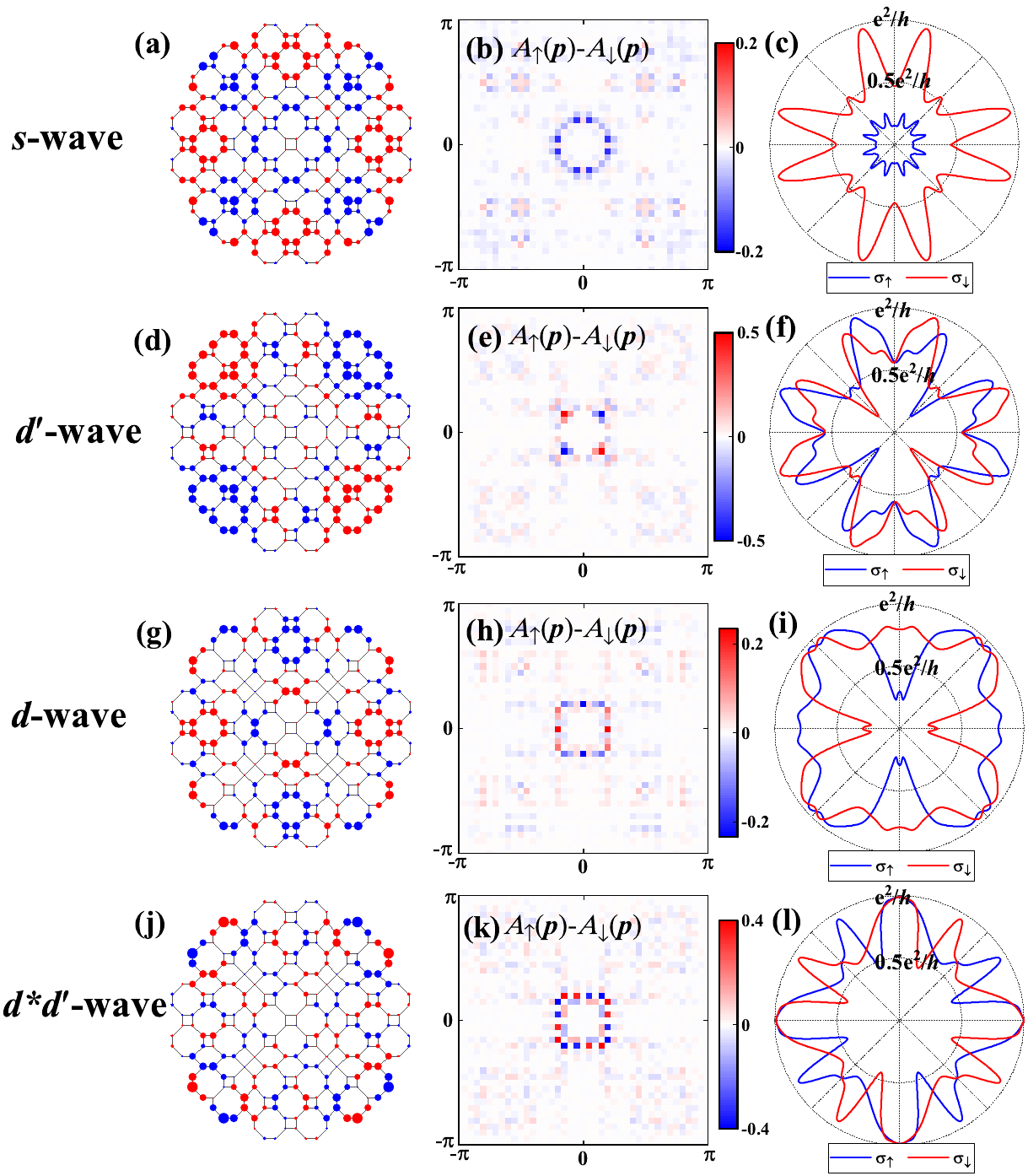}
    \caption{Magnetic states in the Thue-Morse QC with $D_4$ symetry and their properties. (a) The spatial distribution of local magnetic order $\Delta_i$ with $s$-wave symmetry in real space. The color and area of the dots represent the sign and relative magnitude of the magnetic order, respectively. (b) The spin-resolved spectral density $\mathcal{A}_{\uparrow}(\bm{p})-\mathcal{A}_{\downarrow}(\bm{p})$ as a function of momentum $\bm{p}$. (c) Spin conductances $\sigma_{\uparrow}$ and $\sigma_{\downarrow}$ as functions of the angle $\phi$ with $R=3.5$. The value of $\sigma$ is represented by the radial distance. (d-f), (g-i) and (j-l) are similar to (a-c), except that the symmetry of the shown state is $d^{\prime}$-, $d$- and $d*d^{\prime}$-wave, respectively. The chemical potential is fixed at $\mu = -1.28$ in all panels.} \label{fig_SD4}
\end{figure}

In Fig. \ref{fig_SD4}, we also show some results of the magnetic states in the Thue-Morse QC with $D_4$ symmetry and their properties. $D_4$ group has four 1D IRRPs, $A_1$, $A_2$, $B_1$ and $B_2$, corresponding to $s$-wave, 
$d*d^{\prime}$-wave, $d$-wave, and $d^{\prime}$-wave
phases, respectively. Just like in the prior case for $D_8$, the $s$-wave phase is FM state, and other phases are AM states.

An $s$-wave state and its properties are shown in Figs. \ref{fig_SD4} (a-c). 
The magnetic pattern does not change under any symmetry operation, as shown in Fig. \ref{fig_SD4} (a), indicating that this state is FM. 
Both spin-resolved spectrum difference $\mathcal{A}_{\uparrow}-\mathcal{A}_{\downarrow}$ (Fig. \ref{fig_SD4} (b)) and spin conductance difference $\sigma_{\uparrow}-\sigma_{\downarrow}$ are non-zero and do not change under any symmetry operation, also indicating that this state is FM. 
Table I in the main text is verified by this example. 

A $d^{\prime}$-wave state, a $d$-wave state and a $d*d^{\prime}$-wave state, and their properties are shown in Figs. \ref{fig_SD4} (d-f), (g-i) and (j-l), respectively. 
These states are $\mathcal{P}$ even and thus $\mathcal{PT}$ odd, but have zero total magnetism protected by mirror reflection or $c_4^1$ rotation, as shown in Figs. \ref{fig_SD4} (d), (g) and (j), respectively. 
The spin-resolved spectrum $\mathcal{A}_{\uparrow}-\mathcal{A}_{\downarrow}$ (Figs. \ref{fig_SD4} (e), (h) and (k)) and spin conductance difference $\sigma_{\uparrow}-\sigma_{\downarrow}$ (Figs. \ref{fig_SD4} (f), (i) and (l)) are non-zero and have the same symmetry as the corresponding state, indicating that all these states are AM. Table I in the main text is verified by all these examples. 

\begin{figure}[htbp]
    \centering
    \includegraphics[width=0.8\linewidth]{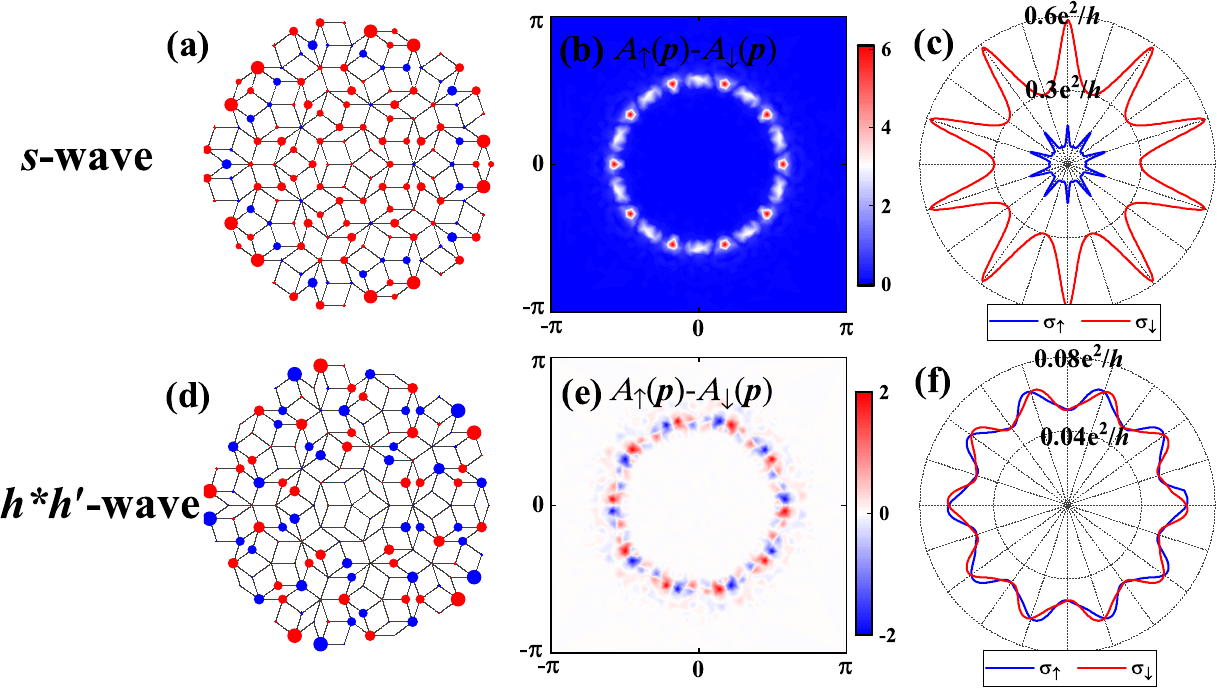}
    \caption{Magnetic states in the Penrose QC with $D_5$ symmetry and their properties. (a) The spatial distribution of local magnetic order $\Delta_i$ with $s$-wave symmetry in real space. The color and area of the dots represent the sign and relative magnitude of the magnetic order, respectively. (b) The spin-resolved spectral density $\mathcal{A}_{\uparrow}(\bm{p})-\mathcal{A}_{\downarrow}(\bm{p})$ as a function of momentum $\bm{p}$. (c) Spin conductances $\sigma_{\uparrow}$ and $\sigma_{\downarrow}$ as functions of the angle $\phi$ with $R=5$. The value of $\sigma$ is represented by the radial distance. (d-f) are similar to (a-c), except that the symmetry of the shown state is  $h*h^{\prime}$-wave and $R=3.5$ in (f). The chemical potential is fixed at $\mu = -0.155$ in all panels.} \label{fig_SD5}
    
\end{figure}

The results for $n=5$ are shown in Fig. \ref{fig_SD5}. 
The magnetic pattern for a typical $s$-wave state, corresponding to the identity IRRP $A_1$, and the spin-resolved spectrum difference $\mathcal{A}_{\uparrow}(\bm{p})-\mathcal{A}_{\downarrow}(\bm{p})$ and spin conductance $\sigma_{\uparrow}$ and $\sigma_{\downarrow}$ for this state are shown in Figs. \ref{fig_SD5} (a-c), respectively. As shown in Fig. \ref{fig_SD5} (a), this state does not change under any symmetry operation in the point group, leading to no restriction to the total magnetism, thus this state is FM. The spin-resolved spectrum difference $\mathcal{A}_{\uparrow}(\bm{p})-\mathcal{A}_{\downarrow}(\bm{p})$ (Fig. \ref{fig_SD5} (b)) and spin conductance difference $\sigma_{\uparrow}-\sigma_{\downarrow}$ are non-zero and also do not change under any symmetry operation, also indicating that the state is FM. These results verify Table I in the main text. 

The magnetic pattern for a typical $h*h^{\prime}$-wave state, corresponding to IRRP $A_2$, and the spin-resolved spectrum difference $\mathcal{A}_{\uparrow}(\bm{p})-\mathcal{A}_{\downarrow}(\bm{p})$ and spin conductance $\sigma_{\uparrow}$ and $\sigma_{\downarrow}$ for this state are shown in Figs. \ref{fig_SD5} (d-f), respectively. As shown in Fig. \ref{fig_SD5} (d), this state does not have $\mathcal{P}$ symmetry and thus does not have $\mathcal{P}\mathcal{T}$ symmetry. However, this state has zero total magnetism protected by mirror reflection. The spin-resolved spectrum difference $\mathcal{A}_{\uparrow}(\bm{p})-\mathcal{A}_{\downarrow}(\bm{p})$ (Fig. \ref{fig_SD5} (e)) and spin conductance difference $\sigma_{\uparrow}-\sigma_{\downarrow}$ (Fig. \ref{fig_SD5} (f)) are non-zero and have the same symmetry as the corresponding state, indicating that this state is AM. These results verify Table I in the main text. 

\twocolumngrid


\bibliography{references}

\end{document}